\documentclass[a4paper]{mem}
\usepackage{natbib}
\usepackage{graphicx}
\usepackage[a4paper]{hyperref}
\idline{0}{1}
\begin{document}
   \title{Spatial distributions of the 
REFLEX-DXL galaxy clusters at $z \sim 0.3$ observed by
XMM-Newton
\thanks{This work is based on observations
made with the XMM-Newton, an ESA science mission with 
instruments and contributions directly funded by
ESA member states and the USA (NASA).}
}

   \author{Y.-Y. Zhang\inst{1}, A. Finoguenov\inst{1}, 
H. B\"ohringer\inst{1},
Y. Ikebe\inst{1,2}, K. Matsushita\inst{1,3}, 
P. Schuecker\inst{1},L. Guzzo\inst{4} and
C. A. Collins\inst{5}
}

   \offprints{Y.-Y. Zhang,\\
email: yyzhang@mpe.mpg.de }

   \institute{Max-Planck-Institut f\"ur extraterrestrische Physik, Garching, Germany; 
\and Joint Center for Astrophysics, University of Maryland, Baltimore, USA; 
\and Tokyo University of Science, Tokyo, Japan;
\and INAF - Osservatorio Astronomico di Brera, Merate/Milano, Italy;
\and Liverpool John Moores University, Liverpool, U.K.}

   \abstract{
We present XMM-Newton results on the spatially resolved
temperature profiles of eight massive galaxy clusters of a
volume-limited sample at redshifts $z\sim0.3$ (REFLEX-DXL sample) and an
additional luminous cluster at $z=0.2578$,
selected from the REFLEX survey.   
Useful temperature measurements
could be performed out to radii with overdensity 500 ($r_{500}$).
The scaled temperature distributions show good similarities. 
We discovered diversities in the temperature
gradients at the outer cluster radii with examples of both flat and
strongly decreasing profiles which call for different physical 
interpretations. 
We found an indication of the 'warm-hot' gas existing 
in or around the hot clusters. 
Using RXCJ0307.0$-$2840 we demonstrate that the
errors on the mass estimates are within
25\% up to $r_{500}$. 

   }
   \authorrunning{Zhang et al.}
   \titlerunning{Spatial distributions of the REFLEX-DXL galaxy clusters}
   \maketitle
%

\section{Introduction}

The most massive clusters are
especially important in tracing large scale structure (LSS) 
evolution since they are expected
to show the largest evolutionary effects. In hierarchical 
modeling the structure of the X-ray emitting
plasma in the most massive clusters is 
essentially determined by gravitational effects
and shock heating. With decreasing cluster mass and intracluster
medium (ICM) temperature, non-gravitational effects play an important
role before and after the shock heating (Voit \& Bryan 2001; Voit et
al. 2002; Zhang \& Wu 2003; Ponman et al. 2003).  Therefore, the most
massive clusters provide the cleanest results in comparing theory with
observations.

\begin{figure}
   \includegraphics
[width=5.6cm]{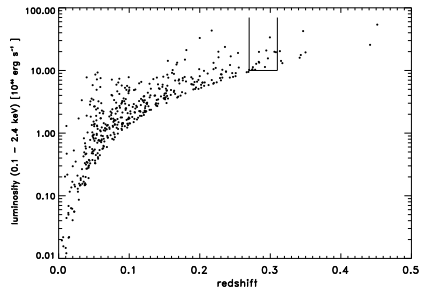}
   \caption{X-ray luminosity-redshift distribution 
of the REFLEX clusters. The box shows the 
selection of the 13 REFLEX-DXL clusters  
(see B\"ohringer et al. 2003).}
\label{f:reflex1}
    \end{figure}

In this project we are analysing an almost volume-complete sample of thirteen
X-ray luminous ($L_{X}~\geq~10^{45}~{\rm erg~s^{-1}}$ 
for 0.1--2.4~keV)
clusters selected from the ROSAT-ESO Flux-Limited X-ray (REFLEX)
galaxy cluster survey (B\"ohringer et al. 2001) in the redshift
interval $z=0.27$ to $0.31$ (see Fig.~\ref{f:reflex1}).
   \begin{figure}
\includegraphics[width=6cm]{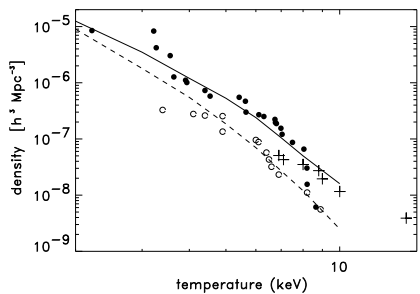}
   \caption{A preliminary REFLEX-DXL temperature 
function (crosses) is compared to the $<z>=0.05$ 
(solid circles) and $<z>=0.38$ 
(open circles) temperature functions from Henry (2000).
See B\"ohringer et al. 2003.}
\label{f:reflex2}
    \end{figure}
There is only a very small correction
to the volume completeness with a well known selection function
for $L_{X}~\geq~10^{45}~{\rm erg~s^{-1}}$ at the higher redshift  
as described in B\"ohringer et al. (2003).
With this REFLEX-DXL (Distant X-ray
Luminous) sample, we obtain reliable
ICM temperatures to measure the cluster masses
based on the high 
resolution observations from XMM-Newton
(Zhang et al 2004).
Since peculiarities in the cluster structure introduce a
scatter in the mass-temperature relation and since in particular
on-going cluster mergers can lead to a temporary increase in the
cluster temperature and X-ray luminosity (Randall et al. 2002), we aim
for a detailed study of the deep XMM-Newton observations described
here. The clusters are also scheduled for a detailed spectroscopic
study of the cluster dynamics with the ESO-VLT-VIMOS instrument.
One prime goal is to study 
the temperature function evolution 
(see Fig.~\ref{f:reflex2}, B\"ohringer et al. 2003) by 
comparing our sample with more
nearby and more distant clusters in literature.
The selection of the REFLEX-DXL sample and its properties are
described in detail in B\"ohringer et al. (2003).
The method is well described in Zhang et al. (2004), which is 
established for a reliable determination of the spatially resolved 
temperature profiles for the REFLEX-DXL clusters. 
XMM-Newton with its superior sensitivity combined with its good
spatial resolution provides the best means for such studies (Arnaud et
al. 2002). Previously, large data sets on cluster temperature profiles
have been compiled from ASCA (e.g. Markevitch et al. 1998; White 2000;
Finoguenov et al. 2001a; Finoguenov et al. 2002; Sanderson et al. 2003)
and BeppoSAX observations (Molendi \& De Grandi 1999; Ettori et
al. 2002).

In this proceeding we contribute the temperature profile measurements,
discuss physics of diversity,
describe an indication for soft excess from warm InterGalactic Medium 
(IGM), and present the mass estimates. We
adopt a flat $\Lambda$CDM cosmology with the density parameter
$\Omega_{\rm m}=0.3$ and the Hubble constant $H_{\rm
0}=70$~km~s$^{-1}$~Mpc$^{-1}$. Error bars correspond to the 68\%
confidence level, unless explicitly stated otherwise.

\section {Temperature distributions}
\label{s:method}

We developed a reliable double background subtraction method for 
the XMM-Newton data reduction,
in which we use the 
XMM-Newton observations of the Chandra Deep Field South (CDFS) 
background and model the difference 
of the target and CDFS backgrounds with the data from the outer 
Field of view (FOV). The details are available in Zhang et al (2004).

Comparing the spectral results,
we have noted that the differences between the global
temperatures of the regions covering radii of 
$0.5<r<4^\prime$ and $r<8^\prime$, respectively,
are caused by systematic differences in the temperature
gradients. For a more detailed study of the temperature profiles we
divide the cluster regions into the five annuli 0--0.5$^\prime$,
0.5--1$^\prime$, 1--2$^\prime$, 2--4$^\prime$, and 4--8$^\prime$. 

We use $>1$~keV band except for RXCJ0658.5$-$5556. We
apply the 2--12~keV band for this high temperature cluster.  The
temperature profile of each cluster is shown 
in Zhang et al. (2004).
We detect the gradients in the spatially resolved temperature profiles
with an accuracy of better than 10 to 20\% in the $r<4^\prime$ region.
The temperatures vary as a function of the radius by a factor of 1.5
to 2. 
To some degree the difference of the central structure 
might reveal the effect of non-gravitational
processes and radiative cooling. 
No significant cooling gas lower than 2 keV is found in the center. 

\begin{figure*}
   \centering
   \includegraphics[angle=-90,width=12.cm]
{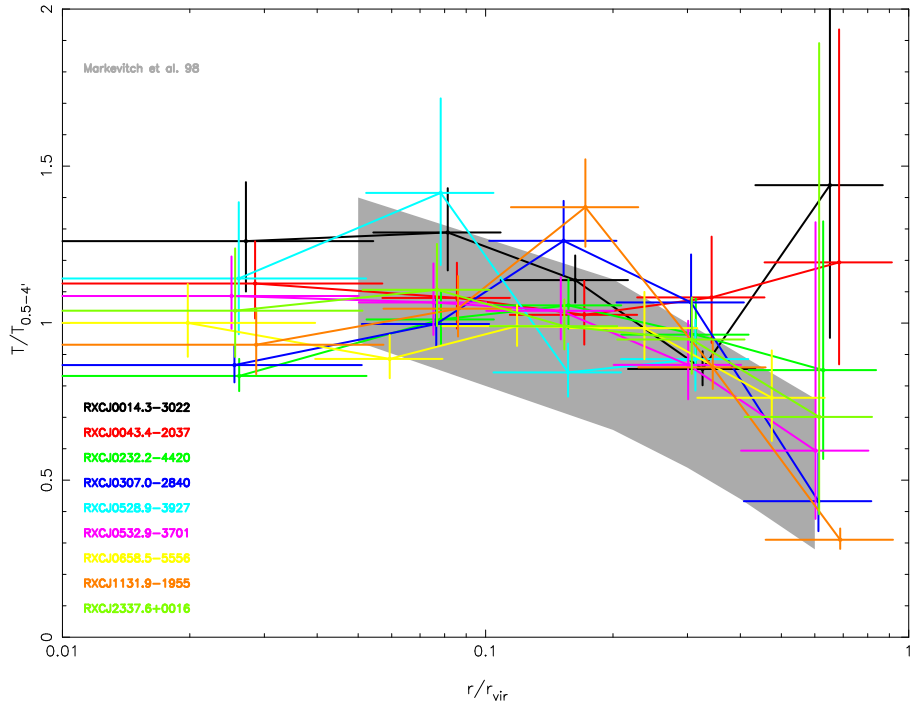}
\caption{Scaled temperature profiles fitted
in the 1--10~keV band, noted
the temperature profile of RXCJ0658.5$-$5556 was 
fitted in the 2--12~keV band.}
\label{f:ktcomp2}%
\end{figure*}

In Fig.~\ref{f:ktcomp2}, we present the scaled temperatures 
of eight REFLEX-DXL clusters together with 
an additional cluster at slightly lower redshift $z=0.2578$.
The radii are scaled by the virial radii obtained from the M-T 
relation in Bryan and Norman (1998). The temperatures are scaled by 
the temperatures of the regions covering radii of $0.5<r<4^\prime$, 
which cover almost the whole X-ray emission regions of 
the clusters.
The grey shadow shows the self-similar temperature profile of  
Markevitch et al. 
(1998). The temperature profiles are on average flattened and
the high accuracy of our data 
shows that the decrease in our temperature 
profiles happens at further out radii. There are also diversities
in the temperature profiles which were not observed before. 

\section{Physical interpretations}
\label{s:physics}

\begin{figure*}
\centering
\includegraphics[width=4cm]{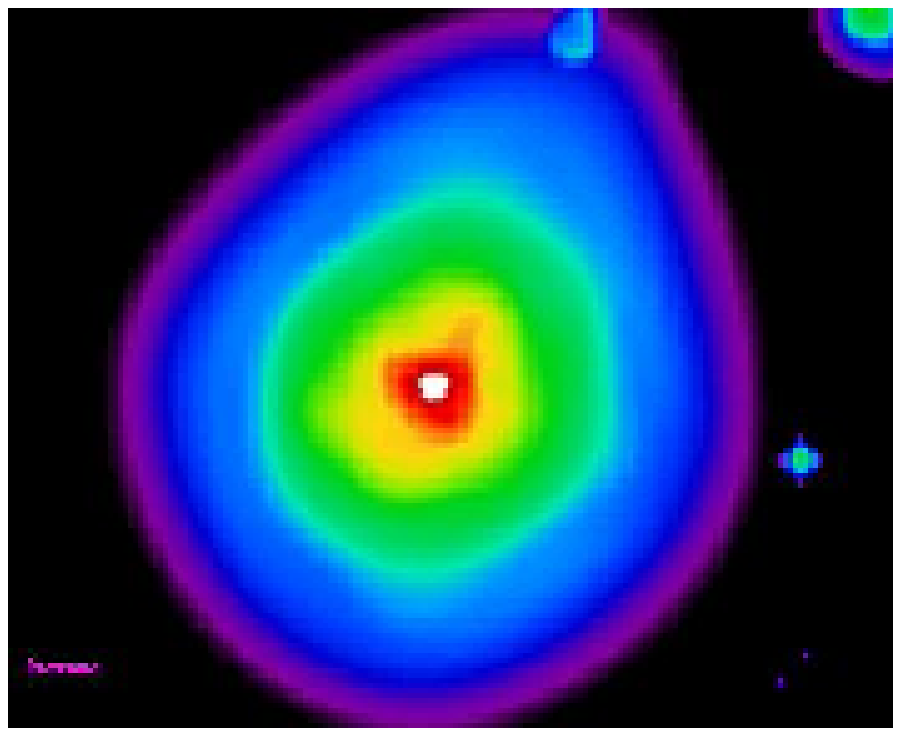}
\includegraphics[width=4cm]{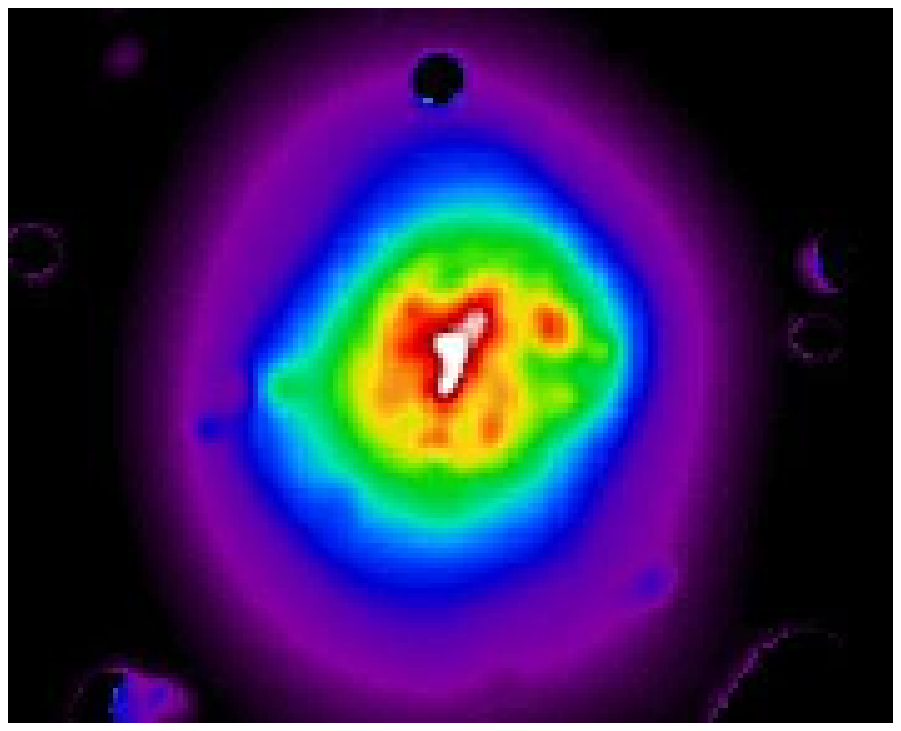}
\includegraphics[width=4cm]{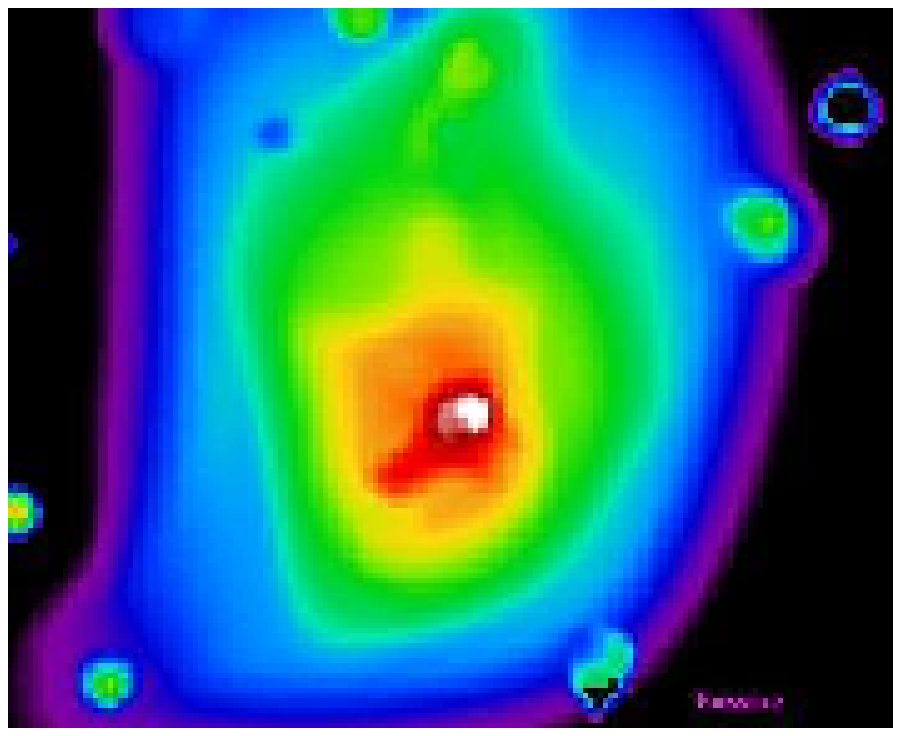}
\caption{ 
Pressure maps 
of RXCJ0307.0$-$2840 (left), RXCJ0658.5$-$5556 (middle) 
and RXCJ1131.9$-$1955 (right). }
\label{f:cl052d}%
\end{figure*}

The variations of the temperature profiles call for different
physical interpretations rather than simple reflections of 
measurement errors (see Fig.~\ref{f:cl052d}). 
Symmetric X-ray 2-D maps and regular temperature profiles suggest 
a relaxed dynamical state, e.g. RXCJ0307.0$-$2840.
On the other hand irregular temperature profiles
like diversities in the temperature gradients, e.g. RXCJ0658.5$-$5556,
suggest mergers and/or significant substructures
in those particular regions. Detailed spectroscopy
of the particular regions
should be performed based on the combination of the 
temperature profile and 2-D map implications. 
Since deprojection dilutes the 
fluctuations in 2-D maps, some clusters have regular 
temperature profiles but very disturbed structures in their 2-D maps,
e.g. RXCJ1131.9$-$1955. 
Chandra and VLT observation are granted to study these 
interesting examples via their galaxy dynamics. 

\section{Soft excess}
\label{s:excess}

\begin{figure}
\centering
\includegraphics[width=3.2cm]
{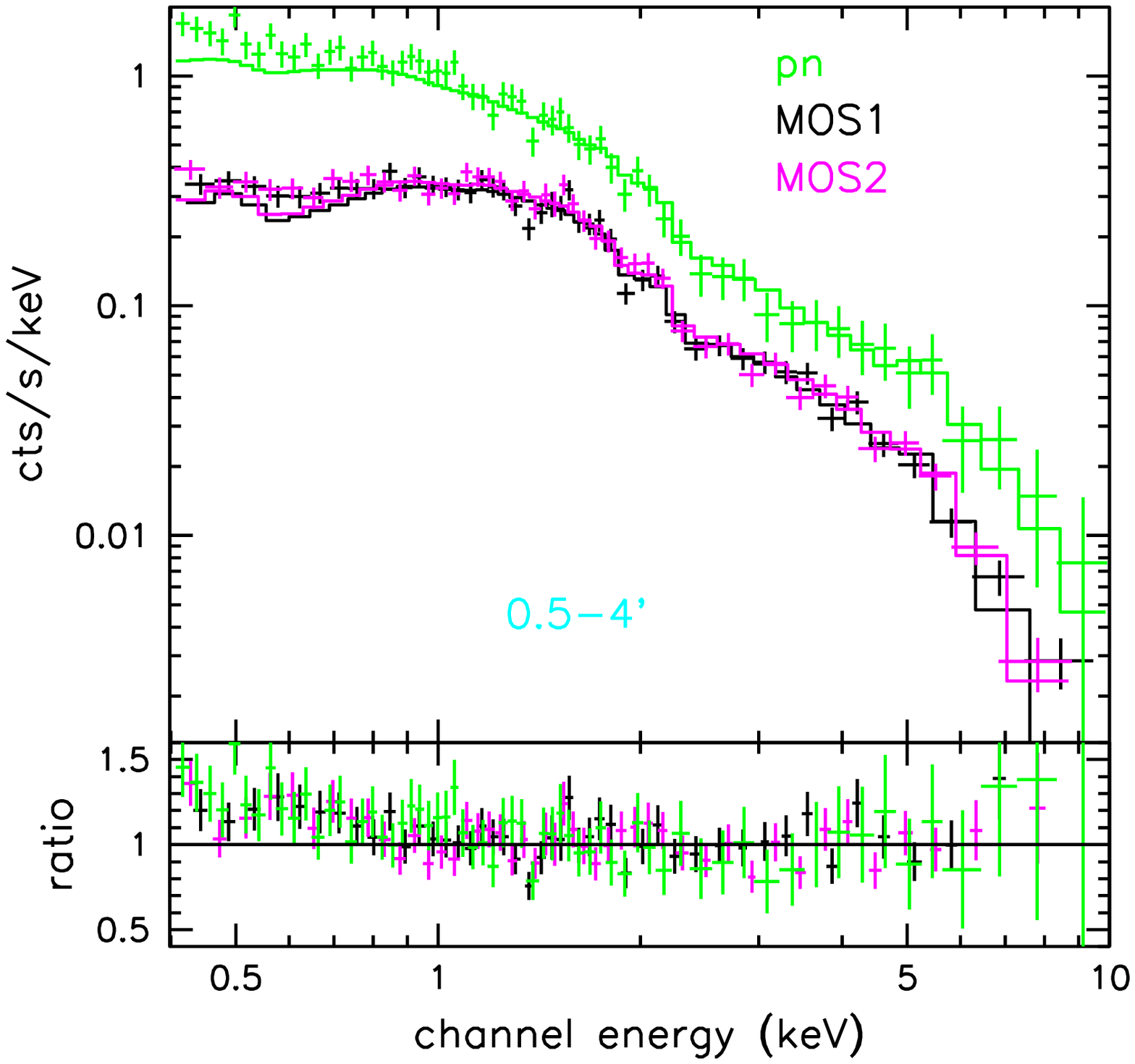}
\includegraphics[width=3.2cm]
{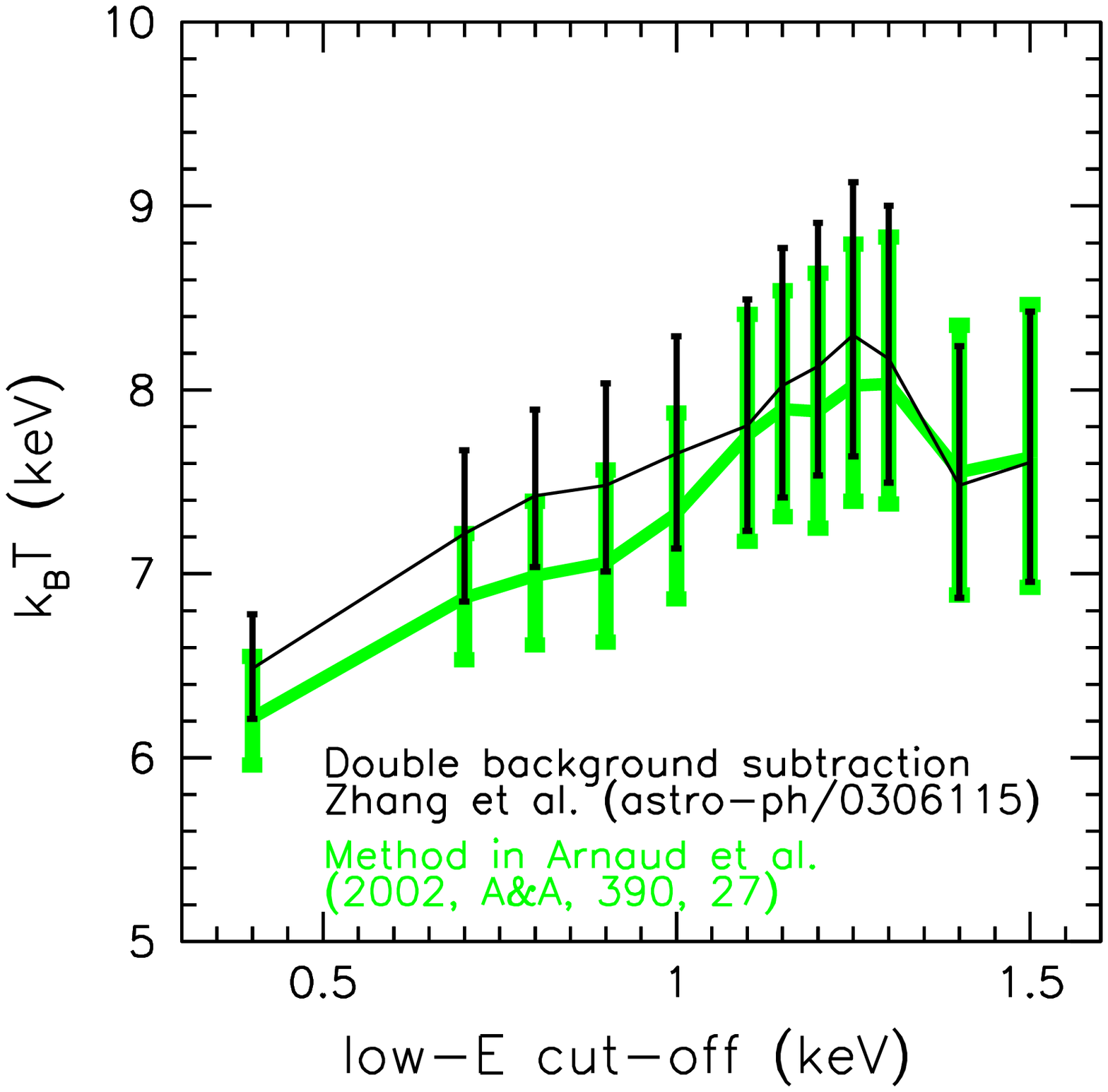}
\caption{Left: Spectra of RXCJ0528.9$-$3927 fitted 
in the 1--10~keV band. Right: Temperature 
measurements versus low-E cut-off.}
\label{f:ktdeviation}%
\end{figure}

We found some inconsistencies in the fitting of the spectra
with one temperature model, depending on the inclusion or exclusion
of the 0.4--1~keV band, which could be an indication of the
presence of a colder component.
Cluster RXCJ1131.9$-$1955 is not
affected at all, RXCJ0014.3$-$3022, RXCJ0307.0$-$2840 and
RXCJ0528.9$-$3927 (see Fig.~\ref{f:ktdeviation}) are 
affected in the center, while RXCJ0043.4$-$2037 and
RXCJ0232.2$-$4420 are affected in the outskirts. Since the instrumental
setup used to observe this sample is the same, it hardly is an
instrumental artifact. 
We notice that the temperature of RXCJ0528.9$-$3927
(also in other clusters) changes 
significantly with the low cut-off of the energy band used in the
fit. We thus performed the X-ray spectral analysis to 
test the energy band dependence
and possible method dependence by comparing the temperature 
measurements versus low energy band (low-E) cut-off 
from two different methods:
the double background subtraction method in Zhang et al. (2004)
and the method applied in Arnaud et al. (2002). 
We found that soft excess exists independent of the method.

The metallicity and 
redshift measurements among the different methods and different 
low-E cut-off vary within 5\%. The results presented in
Fig.~\ref{f:ktdeviation} suggest some influence of the low energy
band on the temperature measurements. 
Thus the global galaxy cluster temperature is underestimated   
including the soft band. The results
obtained in the harder energy band should recover the correct cluster
temperature. Similar phenomena are found for A1413
(Pratt \& Arnaud 2002) using XMM-Newton data, and Coma, A1795 and
A3112 (Nevalainen et al. 2003) based on the comparison of XMM-Newton
and ROSAT PSPC observations. Nevalainen et al. interprete it as a
`warm-hot' intergalactic medium.
The 'warm-hot' gas existing in or around the hot clusters
might be related with the LSS enviroment.

\section{Modeling RXCJ0307.0$-$2840}
\label{s:result}

We use RXCJ0307.0$-$2840 as an illustrative example to demonstrate the
accuracy of measurements of the total gravitating cluster mass and the
gas mass fraction attainable with the XMM-Newton observations of the
REFLEX-DXL-like clusters. Similar analysis of all REFLEX-DXL clusters
is in progress. 

According to the regularity of the photon distribution of 
RXCJ0307.0$-$2840 (see Fig.~\ref{f:cl052d}) 
we assume a radially symmetric gas
distribution. 
We found that the parameterization
\begin{equation}
k_{\rm B}T(r)=\frac{1}{A r^2+Br+C }
\label{e:tprof}
\end{equation}
fits the measured temperature profiles quite well.
The polytropic index ($\sim 1.59$) of the temperature 
distribution in the outskirts implies a convectively stable state there.  
For the electron density distribution
we use the standard $\beta$ model (Michie 1961).
Navarro et al. (1997; NFW) described a universal density profile
for dark matter (DM) from
numerical simulations in hierarchical clustering scenarios.
We assume the intracluster gas to be in hydrostatic equilibrium with
the underlying gravitational potential dominated by DM
component.

\begin{figure}
\centering
\includegraphics[width=3cm]{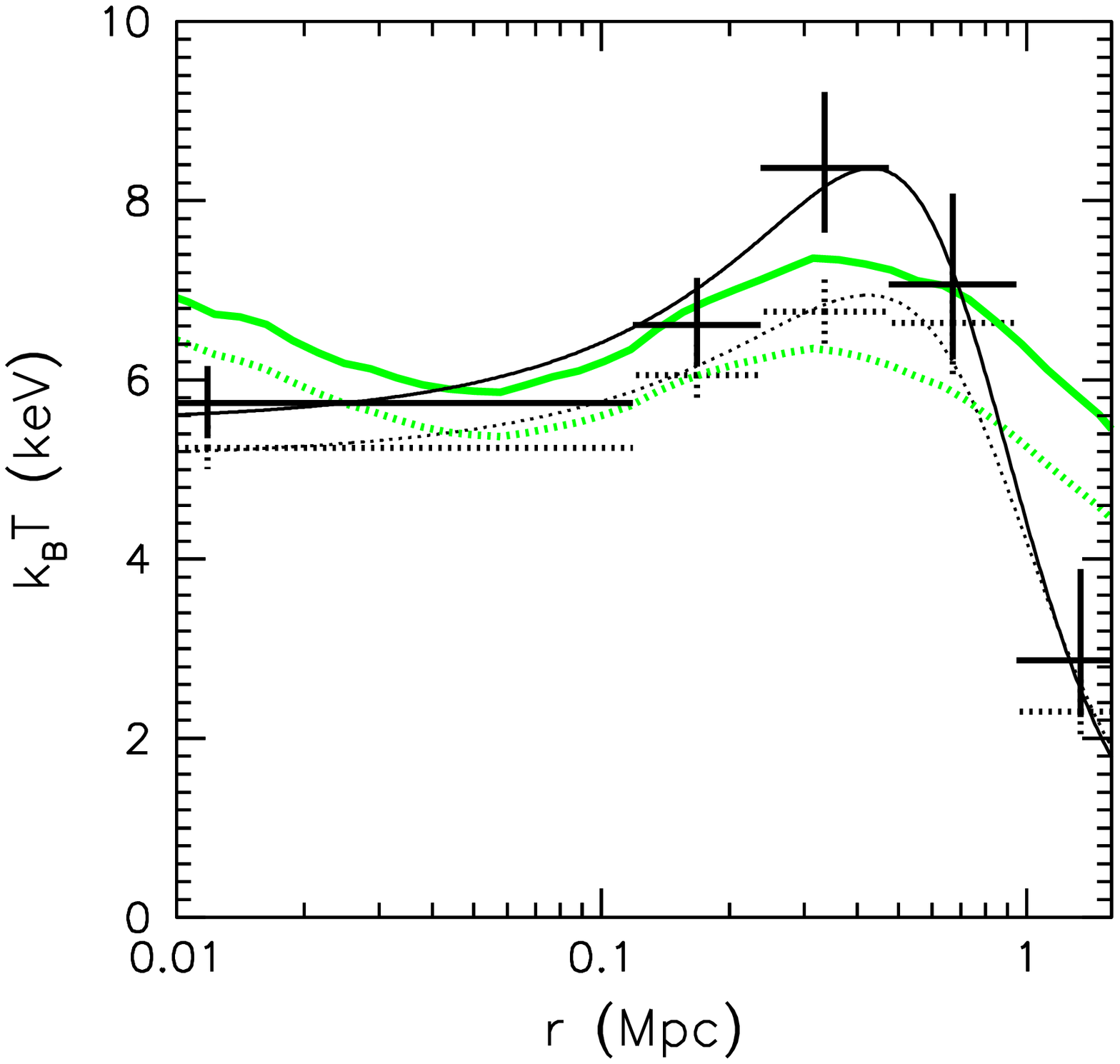}
\includegraphics[width=3cm]{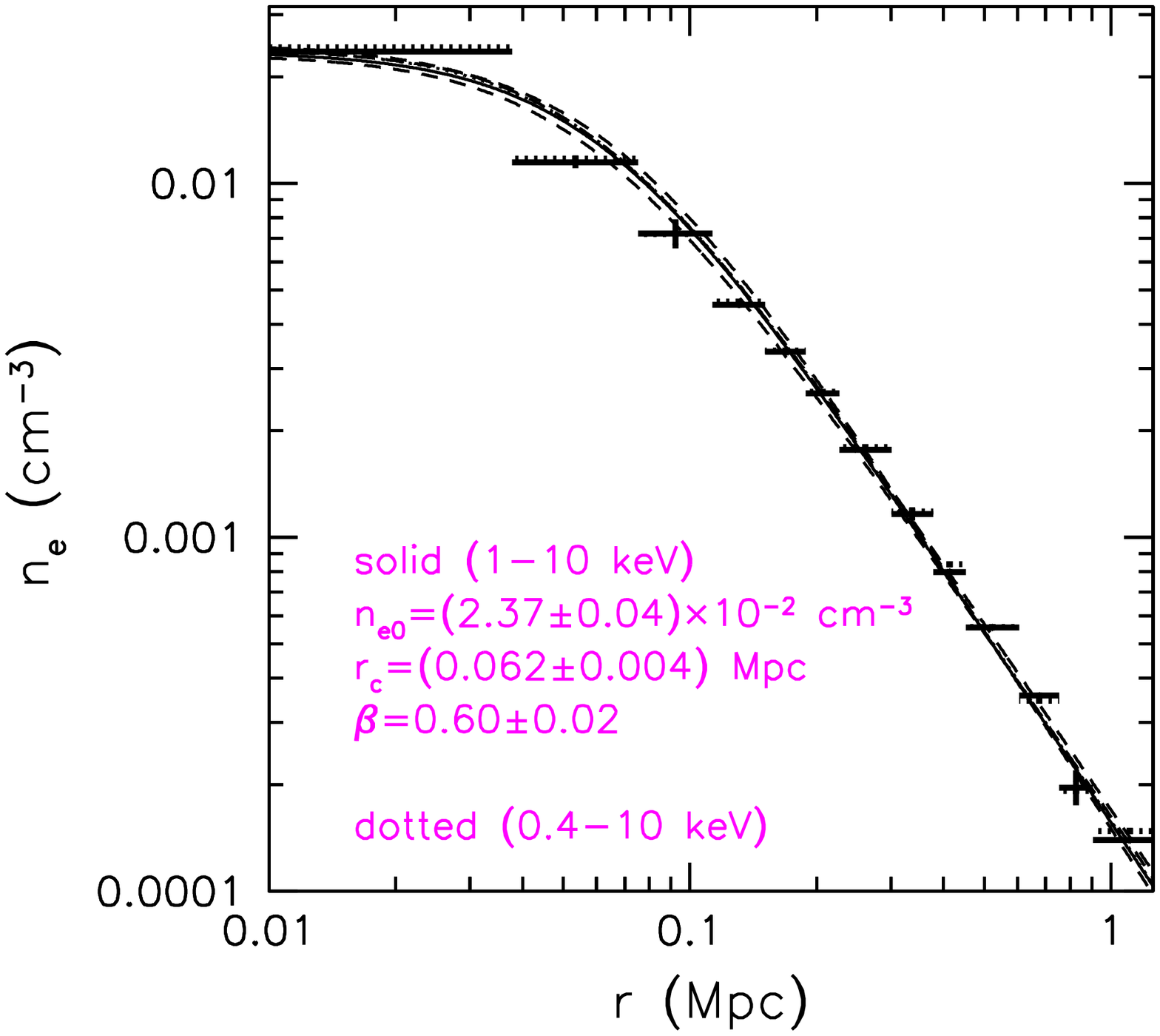}\\
\includegraphics[width=3cm]{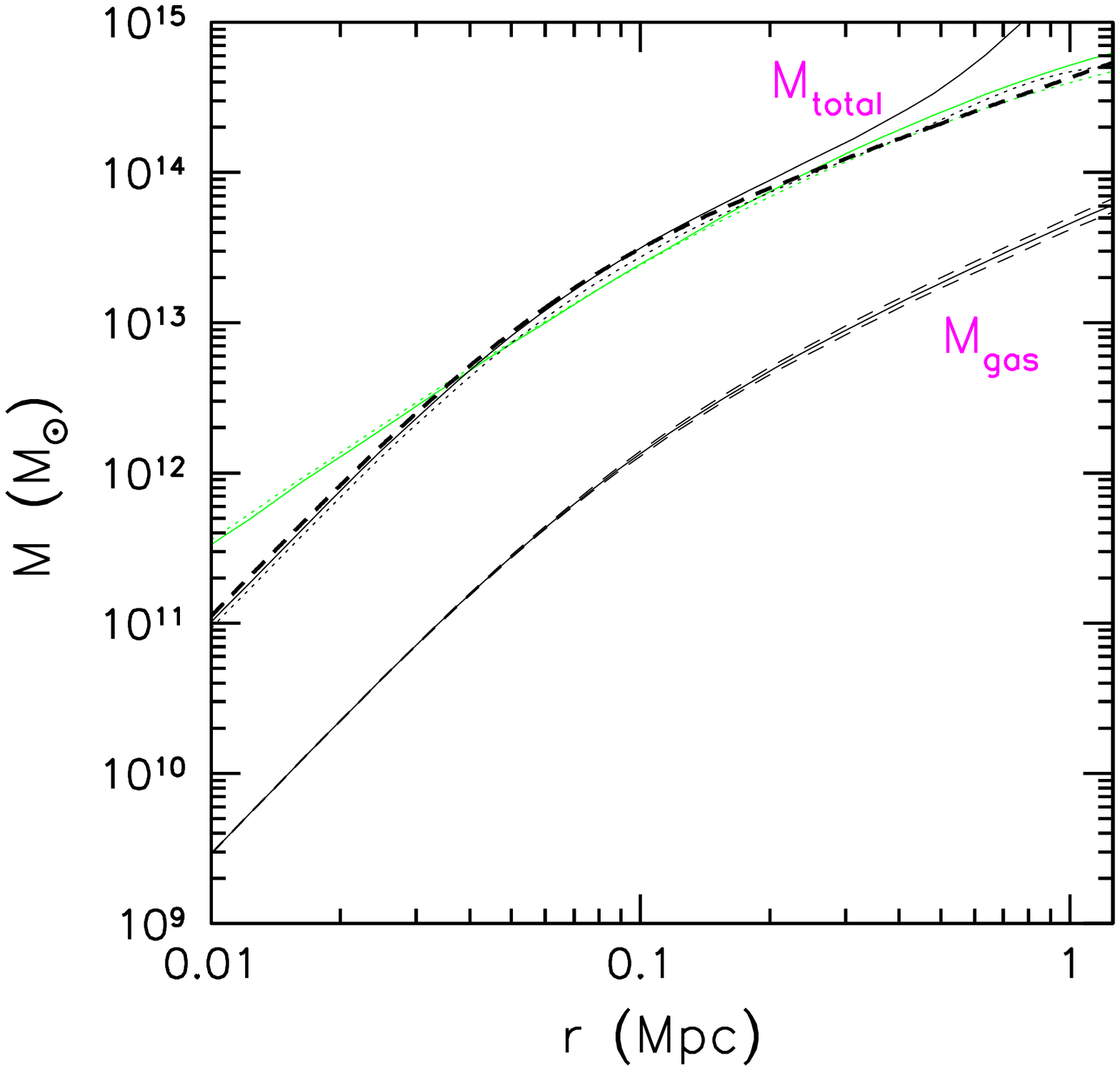}
\includegraphics[width=3cm]{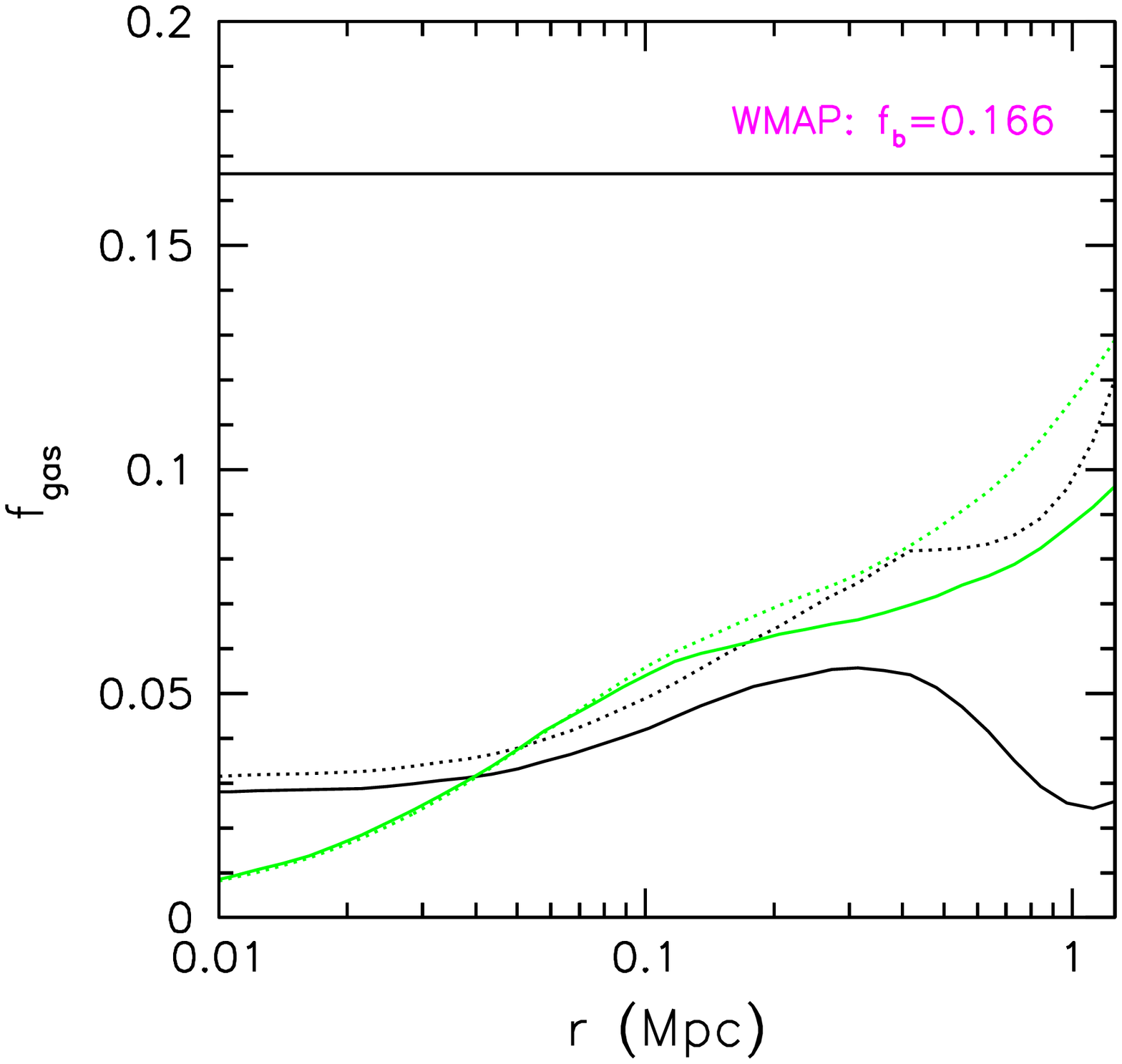}\\
\caption{Spatial 
distributions of the temperature (top left),
electron density (top right), gas mass and 
gravitational mass (bottom left), and gas mass fraction (bottom right) for
RXCJ0307.0$-$2840 using Method I (black) and Method II (green). 
An additional dashed curve of the gravitational mass presents
the result of the isothermal $\beta$ model.}
\label{f:model_4}%
\end{figure}

We thus demonstrate two methods to obtain the modeling of 
the spatial distributions of RXCJ0307.0$-$2840.
One is to combine the $\beta$ model, Eq.(\ref{e:tprof}) and hydrostatic 
equilibrium (here after Method I); 
the other is to combine the $\beta$ model, NFW model 
and hydrostatic 
equilibrium (here after Method II). In Fig~\ref{f:model_4}, 
we present the spatially resolved temperature, electron number density, 
total mass, gas mass, and gas mass fraction  
distributions of RXCJ0307.0$-$2840. 
The uncertainties in the mass estimates
are within 25 percent based on our temperature measurements. 
We found a slightly lower
gas mass fraction comparing the WMAP result.

\section{Summary and Conclusions}
\label{s:conclusion}

We obtain spatially resolved, consistant X-ray temperature profiles 
from three instruments for 9 clusters with an accuracy 
of better than 10--20\% in the 
$<4^\prime$ region. We found 
good similarities in the scaled temperature distributions.
While the high accuracy of our data shows that
the decrease in our data happens at further out radii comparing 
to the self-similar temperature profile of Markevitch et al. (1998).
Temperature structures are easily disturbed by some physical              
processes and are thus complicated.
Differences of the central structures reveal the effect of 
non-gravitational process. No significant cooling gas ($<2$~keV) 
is found in the center. The variations of the temperature 
distributions in the outer region call for different 
formation histories (e.g. Finiguenov et al 2001b). Among the sample, 
six clusters show regular temperature structures, while three do not.
Combination of the regular temperature profiles and symmetric 2-D maps
suggest a relaxed dynamical state. Fluctuations noted in 2-D maps 
might be diluted in the temperature profiles. The 
uncertainty in the temperature measurements is a key point to            
the accurate determinations of mass, gas mass fraction, and thus affects 
the L-T and M-T scaling relations.

\begin{acknowledgements}
      The XMM-Newton project is
supported by the Bundesministerium f\"ur Bildung und Forschung,
Deutschen Zentrum f\"ur Luft und Raumfahrt (BMBF/DLR), the Max-Planck
Society and the Haidenhaim-Stiftung. We acknowledge Jacqueline
Bergeron, PI of the CDFS XMM-Newton observation, Martin
Turner, PI of the RXJ0658.5-5556 XMM-Newton observation,
Michael Freyberg, Ulrich G. Briel, and William R. Forman 
providing useful suggestions.  YYZ
acknowledges receiving the International Max-Planck Research School
Fellowship.  AF acknowledges receiving the Max-Planck-Gesellschaft
Fellowship.  PS acknowledges support under the DLR grant
No.\,50\,OR\,9708\,35. 
\end{acknowledgements}

\bibliographystyle{aa}

\end{document}